\documentclass[aps,prx,amsmath,amssymb,reprint,nofootinbib,superscriptaddress,showkeys,apsrmp4-1]{revtex4-1}
\usepackage[pdftex]{graphicx}
\usepackage{dcolumn, bm, hyphenat}
\usepackage[utf8]{inputenc}
\usepackage[T1]{fontenc}
\usepackage[english]{babel}
\usepackage[mathlines]{lineno}
\begin{document}

\title{The Higher Education Space,\\Connecting Degree Programs from Individuals' Choices}

\author{Cristian Candia}
\email{ccandiav@mit.edu}
\affiliation{Centro de Investigación en Complejidad Social (CICS), Facultad de Gobierno, Universidad del Desarrollo, Santiago, Chile.}
\affiliation{Collective Learning Group, The MIT Media Lab, Massachusetts Institute of Technology}
\affiliation{Department of Networks and Data Science, Central European University, Budapest, Hungary}

\author{Sara Encarnação}
\email{sarenc@fcsh.unl.pt}
\affiliation{Interdisciplinary Centre of Social Sciences (CICS.NOVA, FCSH), Universidade Nova de Lisboa, Lisboa, Portugal}
\affiliation{Applications of Theoretical Physics Group, Porto Salvo, Portugal}

\author{Flávio L. Pinheiro}
\email{fpinheiro@novaims.unl.pt}
\affiliation{Nova Information Management School (NOVA IMS), Universidade Nova de Lisboa, Lisboa, Portugal}
\affiliation{Collective Learning Group, The MIT Media Lab, Massachusetts Institute of Technology}

\begin{abstract}
Is it possible to derive organizing principles of higher education systems from the applicants' choices? Here, we introduce the Higher Education Space (HES) as a way to describe the complex relationship between degree programs. The HES is based on the application of methods from network science to data on the revealed preferences of applicants to the higher education systems of Chile and Portugal. Our work reveals: 1) the existence of a positive assortment of features -- such as gender balance, application scores, unemployment levels, demand-supply ratio, etc -- along the network structure of the HES; 2) that the decaying of the prevalence level of a feature from a focal degree program extends beyond the dyadic relationships captured by the HES; 3) temporal variations in different features do not spillover/propagate throughout the system in the same way; 4) differences in unemployment levels reported among pairs of degree programs are minimized when taking into consideration the connectivity structure of the HES, largely outperforming the differences in matched pairs using traditional similarity measures; and 5) grouping of degree programs based in the network structure of the HES provides an applicant perspective that complements existing classification systems. Our findings support the HES as a multi-dimensional framework that can effectively contribute to the governance of higher education systems.
\end{abstract}
\maketitle

\section{Introduction}
While many factors are known to determine applicants’ choices when entering Higher Education and to contribute to their educational attainment -- examples ranging from the socio-economic background of applicants \cite{archer2005higher,reay2001choices, sewell1967socioeconomic,goyette2006studies} to their gender \cite{barone2011some,hendley2015gender,stoet2018gender}, but also including the expected earnings differentials between education fields \cite{boudarbat2008field,Centola:2007cb}; self-identification and career opportunities \cite{holmegaard2014choose,wilkes2015reasons}; ability beliefs and heterogeneous tastes \cite{van2016study,wiswall2014determinants,trautwein2007epistemological}; political views, and applicants personality \cite{porter2006college} -- little is known on how these factors translate into higher order principles of higher education systems. 

Linking individual actions to higher-order organizational principles of social systems has been a long lasting problem in computational social sciences \cite{schelling2006micromotives, conte2012manifesto, lazer2009life, maroulis2010complex, watts2013computational, jacobson2015education}. Such link plays a key role in our ability to design effective governance instruments and interventions, in that their effectiveness is arguably bounded by our understanding of how elements in a system can affect each other. In the context of higher education \cite{daniel2015big,baker2014educational}, a lack of such knowledge materializes in our inability to answer simple questions, such as, how do changes in the demand of a given degree program spillover throughout the system? Would such variation be observed equally across all degree programs or would we, instead, observe a predictable and structured spillover dynamics? And what should we expect regarding other measurable features?

Here, we propose the \textit{Higher Education Space} (HES) as a way to map the interplay and similarity between degree programs and as an instrument to improve the effectiveness of policy-making in higher education. Similarity between degree programs is measured by proxy from the revealed preferences of applicants when applying to higher education. The emerging structure, the HES, is a network that connects pairs of degree programs according to the likelihood that they co-occur in the applicant’s preferences. Therefore, the HES represents ‘how students, not administrators or faculty, think about the grouping of’ degree programs \cite{baker2018understanding}. This structure contrasts with the state of the art classification, the \emph{International Standard Classification of Education} (ISCED) \cite{schneider2013international,schneider2008international}, based on the similarity of degree programs according to their expected course content. 

Our work briefly presents findings that illustrate the relevance of the HES in different topics and in the context of the Portuguese and Chilean higher education systems. These have a similar and centrally run application process to higher education. However, they also contrast in many socio-economic standards. 

The HES reveals the existence of positive autocorrelations \cite{leenders2002modeling} among features\footnote{Features here correspond to students' aggregated characteristics in a particular degree program.} of degree programs. These features include gender balance, application scores, demand-supply ratio, unemployment level, first-year dropout rate, and mobility. The autocorrelations patterns indicate that features tend to be positively assorted throughout the network structure of the HES, meaning that, if a degree program exhibits a high prevalence of, say, female applicants, then, degree programs up to two/three links away will also show a similar prevalence. Furthermore, while some features (\textit{e.g.}, application scores and demand-supply ratio) also exhibit autocorrelations patterns with respect to temporal variations, others do not (e.g., gender balance).

Results also show that autocorrelations regarding unemployment cannot be explained simply by matching elements with similar features. Indeed, the connectivity structure of a degree program in the HES seemingly plays a determinant role in the reported unemployment levels. In that respect, we observe that connected degree programs tend to have similar unemployment levels, even after controlling for feature-matched, but unconnected, degree programs. Naturally, this finding has implications for applicants, since the choice of certain degree programs might later translate in social and capital costs associated with labor mobility.

This manuscript is organized as follows: in Section~\ref{data} we present a short description of the data used in this study; Section~\ref{Results} presents the results along with a detailed discussion; and, in Section~\ref{conclusions} we present concluding remarks by summing all major contributions of this work and its societal implications.

\section{Data}
\label{data}

The dataset consists of the preferences of applicants to the Portuguese (PHES) and Chilean (CHES) Higher Education Systems. While each preference in the application process corresponds to a pair of institution and degree program, here we limit the analysis to the choices of degree programs only.

Data for the PHES was obtained from the General Directorate for Education and Science Statistics, DGEEC\footnote{From the Portuguese \emph{Direção-Geral de Estatísticas da Educação e Ciência}}, through a collaboration with the Agency for Assessment and Accreditation of Higher Education,  A3ES\footnote{From the Portuguese \emph{Agência de Avaliação e Acreditação do Ensino Superior}}. This dataset includes application records to all public higher education institutions between 2008 and 2015.

Data for the CHES was provided by the Department of Evaluation, Measurement and Educational Record, DEMRE\footnote{From the Spanish \emph{Departamento de Evaluación, Medición y Registro Educacional}}. CHES data covers the period between 2006 and 2017 and includes all 36 institutions that belong to the Rectors’ Council of Chilean Universities, CRUCH\footnote{From the Spanish \emph{Consejo de Rectores de las Universidades Chilenas}}.

A detailed description of the similarities and differences between each higher education system can be found in Appendix \ref{a:datadescription}. More information related to the  data cleaning procedures can be found in Appendix \ref{a:datacleaning}.

\subsection{Descriptive Features of Degree Programs}
\label{s:21}
For each degree program we collect different descriptive features aiming to explore autocorrelation patterns that might explain the organization of the HES. These features are assembled from the aggregated data of applicants (application scores, gender, demand, and geographical origin) or from institutional reports (unemployment levels, supply levels, and first-year dropout rates).

Each feature is standardized by year and across all degree programs that make the Higher Education Space in each country. For instance, the gender balance of each degree program is estimated by i) computing the fraction of female enrolled students in each degree program, these values are standardized by ii) subtracting the average fraction of enrolled female students among all degree programs and iii) dividing by the standard deviation, thus obtaining a Z-score. Standardization of the features  yields not only comparable results across time but also information about the deviations of each feature to the average of the entire system. 

In this work we focus on the analysis of the following features: \textit{Gender Balance} (PHES and CHES), given by the fraction of female applicants in each degree program that actually enrolled at the end of the application process; \textit{Application Scores} (PHES and CHES), given by the average score of applicants that enrolled in a degree program; \textit{Demand-Supply} ratio (PHES and CHES), given by the ratio between the number of applicants that chose a given degree program as their first choice\footnote{The first option is often considered to reveal the true preference of an applicant.} by the number of open positions in that same degree program. This normalization ensures that demand is corrected for size effects (\textit{i.e.}, cases in which the sheer size of supply can drive demand). This indicator is similar, in spirit, to the "strength index" \cite{portela2008evaluating} sometimes computed to quantify institutions ability to fill the available offer from the first options of applicants; \textit{Geographical Mobility} (CHES only), given by the distance by car, in km, between the candidate's city of origin and the location of the main campus of the institution of enrollment; \textit{Unemployment Level} (PHES only), compiled and reported by institutions, and finally the \textit{first year dropout} rate (PHES only), given by proxy from the enrollment situation of applicants at the end of the first year. Data on both Unemployment and first-year dropout levels are publicly available at \emph{http://infocursos.mec.pt}.

\section{Results and Discussion}
\label{Results}
\subsection{The Higher Education Space}
\label{s:31}

\begin{figure*}[!t]
 \centering 
  \includegraphics[width=0.83\textwidth]{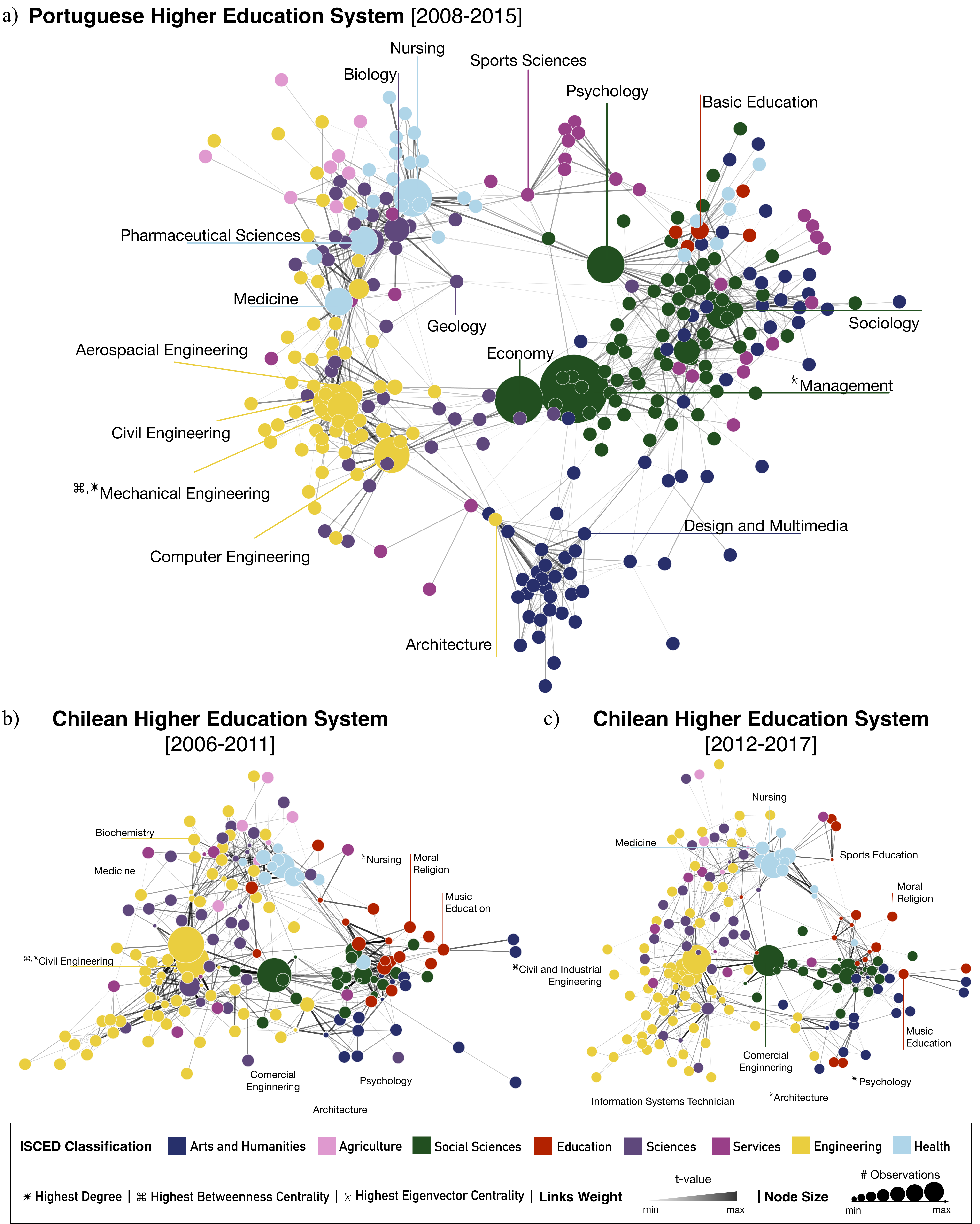}
\caption{The \emph{Higher Education Space} for a) Portugal, from 2008 to 2015 and for Chile, for b) 2006 to 2011  and c) 2012 to 2017. Following the 2012' addition of 9 new universities, the Chilean HE system was divided into two distinct networks (see Appendix \ref{a:datadescription}). The color of the nodes identifies the Education Field of the degree program according to the first level of the International Standard Classification of Education (ISCED). Sizes of nodes illustrate the relative number of observations in relation to other degree programs in the same network, sizes of nodes are not comparable among different networks.}
 \label{figure1}
\end{figure*}

The Higher Education Space (HES) is estimated by identifying which pairs of degree programs exhibit a statistically significant co-occurrence in the applicants' preference list \cite{ronen2014links, hidalgo2009dynamic}. To that end, we start by counting the number of times a pair of different degree programs co-occurs (see Appendix \ref{a:networks} for more details), and control the number of observed co-occurrences by the expected number of occurrences from random chance, based on the total number of observations of each degree program. 

Hence, we start by computing the $\phi$-correlation index $\phi_{ij}$ between pairs of degree programs. This can be achieved by taking a pair of options, $i$ and $j$, and compute:
\begin{equation}
    \phi_{ij}=\frac{M_{ij}N-M_iM_j}{\sqrt{M_i(N-M_i)(N-M_j)}},
\end{equation}
where $M_{ij}$ represents the number of co-occurrences of option $i$ and $j$ in the preferences of a candidate and $M_i$ is the total number of observations of option $i$ ($M_i=\sum_i M_{ij}$).  We discard all negatively correlated relationships, since this implies that such connections appear less than we would expect by random chance.  Moreover, since the magnitude of observations varies across different degree programs, we use a t-test to infer whether the positive correlations are significantly distinguishable from zero. To that end, we compute:
\begin{equation}
    t_{ij}=\phi_{ij}\frac{\sqrt{D-2}}{\sqrt{1-\phi_{ij}^2}},
\end{equation}
where $D-2$ represents the degrees of freedom (here we take a conservative approach and take $D=\text{max}(M_i,M_j)$). All relationships with a $p$-value greater than $0.05$ are discarded as well as all the nodes that are not connected to the giant component.

Figure \ref{figure1} shows graphically the network structures of the HES for Portugal and Chile. Nodes represent degree programs and are colored according to the first level of the ISCED classification, which groups degree programs in one of nine education fields, namely: Arts and Humanities (dark blue), Social Sciences (dark green), Sciences (dark purple), Engineering (dark Yellow), Agriculture (pink), Education (red), Services (light purple), and Health (light blue). The size of the nodes is proportional to the number of observations. Links connect pairs of degree programs with a statistically significant co-occurrence  pattern and thickness is proportional to the t-value associated to the $\phi$-correlation.

The PHES network (Fig. \ref{figure1}a) results from all application preferences between 2008 and 2015, since no major and significant changes occurred in the system during that time interval. By contrast, the CHES network analysis is divided into two periods, due to the 2012’s addition of nine new universities (see Appendix \ref{a:datadescription}). As such, the first period (Fig. \ref{figure1}b) considers all applications between 2006 and 2011, and the second period (Fig. \ref{figure1}c) takes all applications between 2012 and 2017. 

\begin{figure*}[!t]
 \centering 
  \includegraphics[width=.9\textwidth]{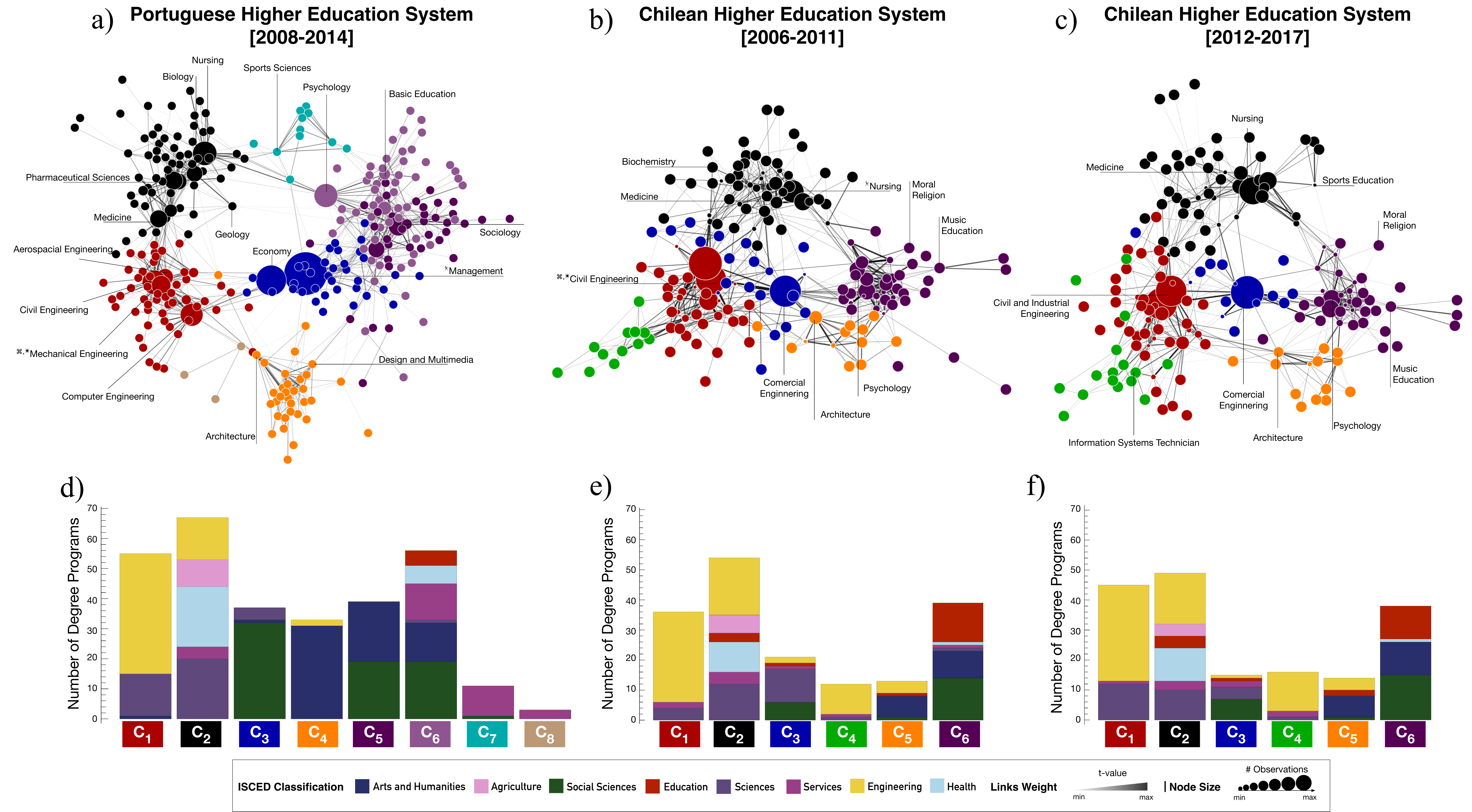}
\caption{Communities of the Higher Education Space in Portugal (a) and Chile (b and c). Nodes are colored according to the community they belong to (see main text for details). Panels (d) to (f) show the composition of each community in accordance with the first level of the ISCED classification. }
 \label{paper2_figure5}
\end{figure*}

The PHES and CHES networks are sparse (between 2$\%$ and 5$\%$ of the maximum number of relationships possible) and highly clustered (clustering coefficient measures between $0.48$ and $0.51$). The high clustering coefficient invites the use of network science methods (\textit{e.g.}, modularity-based network partition algorithms) to derive a classification/grouping of degree programs (see Figure~\ref{paper2_figure5} and related discussion bellow). Each network exhibits a diameter between 6 and 7 links, and an average path length (APL) between $3.06$ and $3.61$. Both CHES networks have fewer nodes than the PHES network ($177$ and $175$ against $301$) but relatively similar connectivity per degree program -- $9.50$ and $8.18$ against $8.15$. There are common topological motifs in all three networks discernible by visual inspection, \emph{viz.} the existence of three main clusters: one dominated by degree programs in Engineering; a second one that involves degree programs in Biology, Sciences, and Health; and a third with a strong representation of degree programs in Arts and Humanities, and Social Sciences.

Overall, the HES space is characterized by a doughnut-shaped structure with a few degree programs occupying a central region connecting opposite sides of the network. This topology is not new and similar networks were obtained when mapping science and research areas \cite{bollen2009clickstream, guevara2016research}. Nonetheless, the above common motifs can have relevant implications for higher education policy development. For example, the centric role of Economics and Management (Commercial Engineering in Chile) connecting the Engineering, Arts and Humanities and Social Sciences clusters might hint to potential trans-disciplinary crossings when designing future changes in the system \cite{iii1997using, shipman1996creating,waugh2016meta}.

As mentioned above, the high clustering levels in all three networks invite for a classification/grouping of degree programs based on the network structure of the HES. Figure~\ref{paper2_figure5} shows the best partitions obtained using the Louvain algorithm \cite{blondel2008fast}, where nodes of the PHES (a) and CHES (c, and d) are colored according to the partition they belong\footnote{To estimate the best partition we have run the Louvain algorithm $10^3$ independent times and selected the partition that resulted in the highest modularity.}. The best PHES partition has a modularity of $0.72$ and explains $86\%$ of the intra-group connectivity. When compared with the ISCED classification, these values correspond to an improvement of $33\%$ in modularity and of $23\%$ in intra-group connectivity. Likewise, the best partition of the CHES networks exhibits a modularity of $0.62$ (2006/11) and $0.63$ (2012/17), explaining $85\%$  (2006/11) and $86\%$ (2012/17) of the intra-group connectivity with an improvement of $5\%$(2006/11) and $9\%$ (2012/17), over the ISCED classification, see Figure~\ref{paper2_figure5}.

Figure \ref{paper2_figure5}d-f shows the composition of each HES group according to the ISCED classification of its constituents. Colors among similar groups ($C_1$ to $C_8$) of different HES are kept consistent to ease the comparison. Groups of similar color match groups located in similar regions of the PHES and CHES. For example, group 1 ($C_1$) in PHES is composed of 11 degree programs from the Science Education Field, 1 degree program from Services and 40 degree programs from Engineering. A similar composition is found in $C_1$ of CHES, and for all other comparable groups ($C_1$ to $C_6$).

As expected, there are differences and similarities among the three HES. Firstly, the number of communities differs between the PHES (8) and the CHES (6) which might be explained simply by the size of each network (see Appendix \ref{a:datadescription} for more details about each system). Secondly, the organization of the CHES network seems to have changed in the second time period, becoming more similar to the PHES network. This conjecture is backed-up by visual inspection only and needs future validation, but raises interesting questions: 1) does globalization of higher education \cite{hazelkorn2015rankings, machin2017paying, killick2014developing,yelland_2011} leads different HES to evolve towards similar structures? and 2) since these structures are based on applicants' choices, are they adapting quickly to societal transformations and is policy on higher education able to follow suit?

\subsection{Feature Assortment in the Higher Education Space}
\label{s:32}
\begin{figure*}[!t]
 	\centering 
	\includegraphics[width=\textwidth]{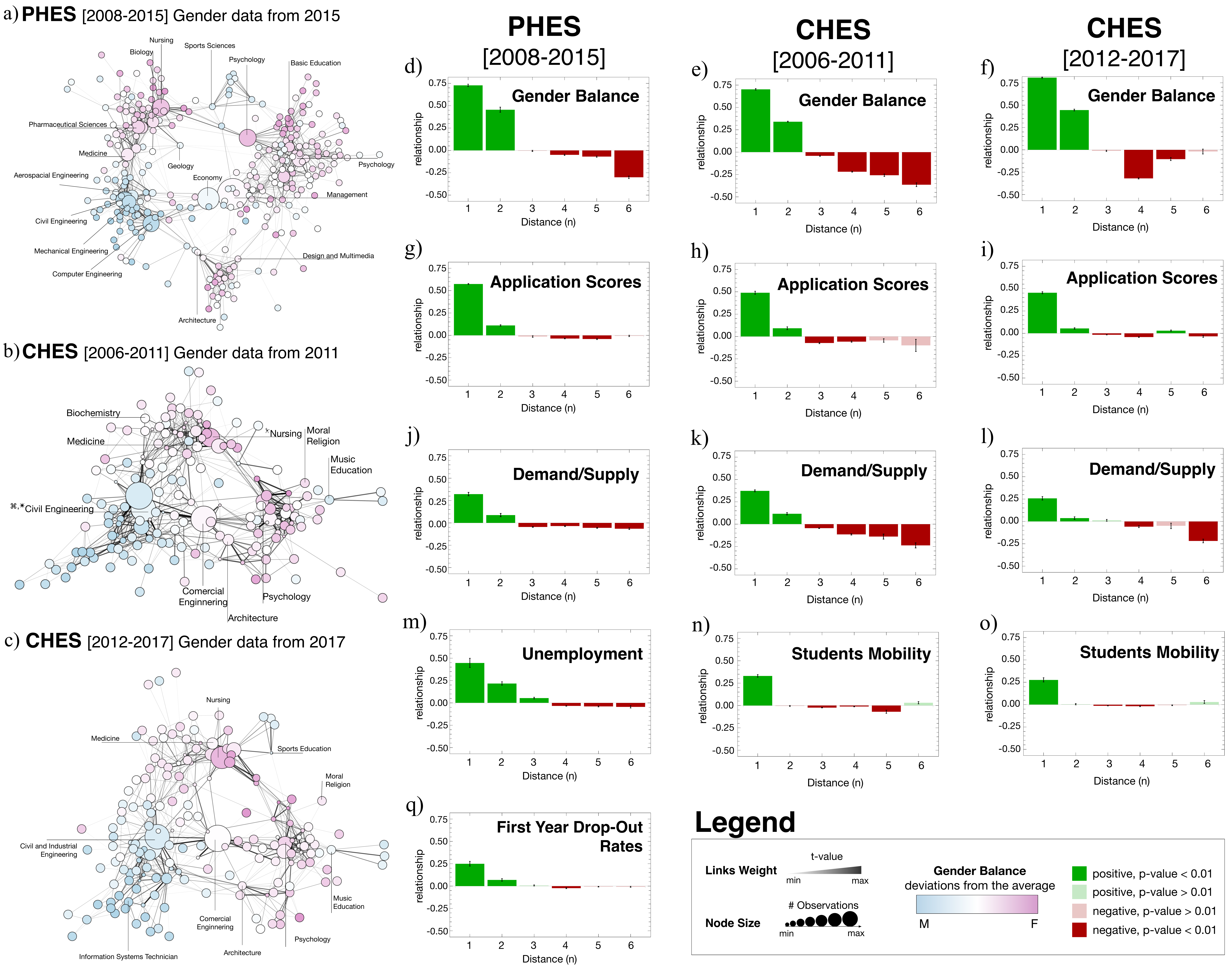}
	\caption{To illustrate the positive assortment of features along the Higher Education Spaces of Portugal and Chile, nodes have been colored according to the gender balance of enrolled students  during the 2015 applications in Portugal (a), 2011 (b) and 2017 (c) in Chile. Panels d--q show the autocorrelations between the aggregated characteristic of pairs of degree program separated by n links in the Higher Education Networks of Portugal (d, g, j, m, and q) and Chile (e,f,h,i,k,l,n, and o). The characteristics under analysis correspond to the gender balance (d--f), application scores (g--i), demand normalized by supply (j--l), unemployment levels (m), students mobility (n--o), and retention rates (q).
	}
	\label{figure2}
\end{figure*}

The \textit{Higher Education Space} (HES) is estimated uniquely based on the applicants' choices and completely nescient about particular features that characterize each degree program. Thus, the emergence of three coherent and similar networks, in two different countries and for different time periods, naturally leads to the question of what explains the emergence of these same structures? The answer likely lies in a multiplicity of factors, some of which we briefly explore here by 
matching the HES network structures with available data on descriptive features of degree programs -- e.g. gender balance or unemployment levels (\emph{c.f.} Section~\ref{s:21}).
It is important to keep in mind that other factors involved in the applicants' choices can certainly help to explain the structure of the HES. However, due to data limitations and the scope of this manuscript such exploration is left for future work. 

Figure \ref{figure2}a-c shows the PHES (a) and CHES (b and c) where each degree program is colored according to the Gender Balance in 2015 (Fig. \ref{figure2}a), 2011 (Fig. \ref{figure2}b), and 2017 (Fig. \ref{figure2}c). Pink (blue) tones identify an above average representation of female (male) applicants. The distribution of Gender prevalence among degree programs is not random or uniform but, in fact, it is clustered, resulting in the predominance of one gender over the other in particular regions of the HES. 
Similar patterns are observed for all other features such as application scores, unemployment levels, demand-supply ratio, mobility, and first-year dropout rates (see Appendix \ref{a:features}).  

Figure \ref{figure2}d-q explores, quantitatively, these clustering patterns (i.e., positive assortment) over the HES. To that end, we compute, for each feature, the autocorrelations between pairs of degree programs at different distances in the HES network (\textit{i.e.}, measured by the minimum number n of  links that form a path from one degree program to the other). Bars represent the autocorrelation averaged over all observation years, and error bars the standard error in the estimation of the coefficients. For example, an autocorrelation of $0.75$ at $n = 1$ for gender dominance, means that degree programs separated by one link exhibit, in average, $75\%$ of the proportion of Female students of a focal degree program. Positive (negative) autocorrelation coefficients are shown in green (red). Bars in light colors indicate an autocorrelation that is not significantly different from zero (failed a t-test with $p > 0.01$).

\begin{figure*}[!t]
    \centering 
    \includegraphics[width=0.9\textwidth]{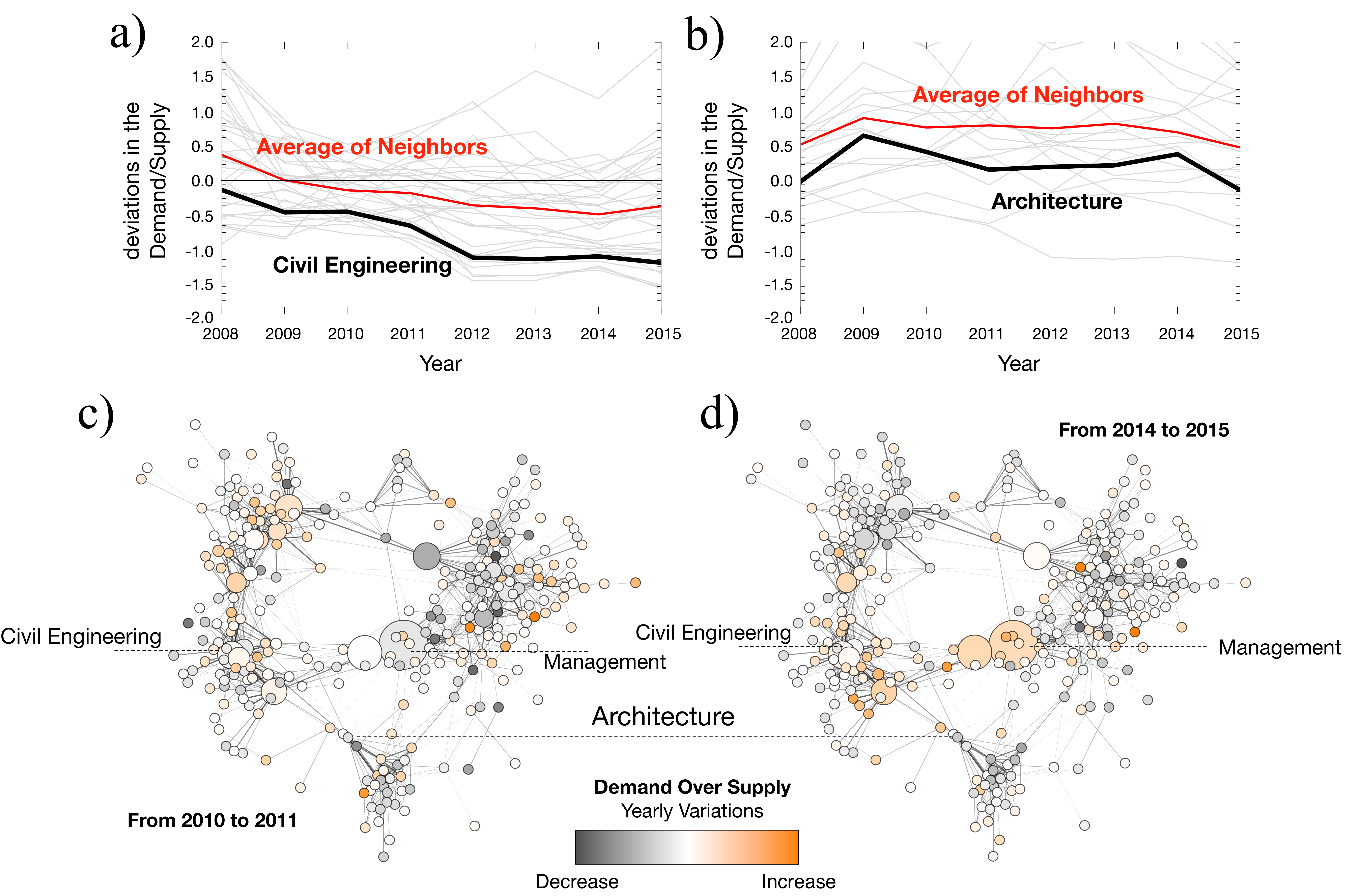}
    \caption{Time variations and spillovers in the Demand-Supply ratio for Civil Engineering (a) and Architecture (b) and their direct neighbours, in the Portuguese Higher Education Network (PHES). The complete PHES network is colored according to the observed Demand-Supply ratio yearly variation between 2010/11 (c) and between 2014/15 (d). Two illustrative examples are shown that capture the variation in demand in a period when the number of applicants was in a clear decline (c) and another when it was growing (d) in the PHES. Orange denotes a positive variation (increase in demand-supply), black a negative variation (decrease in demand-supply).
    }
    \label{figure3}
\end{figure*}

These positive/negative relationships between pairs of degree programs seem to ascertain previous findings \cite{sewell1967socioeconomic,goyette2006studies}, in that some groups of students tend to choose similar preferences based on similar determinants of choice. For example, a positive assortment of in gender balance (Fig.\ref{figure2}d-f) confirms the existence of different preferences from individuals from different gender groups that are revealed in the choices of degree programs, as found in \cite{barone2011some,hendley2015gender,gabay2014gender, stoet2018gender}. But more importantly, and a non-trivial finding of this approach, is to be able to show \emph{How} and \emph{Where} these similarities spread through the network and how neighbouring degree programs (nodes) influence or contaminate each other. In other words, how features spillover throughout the network structure of the HES. Returning to the gender balance example, Figure \ref{figure2}d-f confirms what was already concluded from a visual inspection of the network -- the more female applicants apply to a degree program, the more female applicants are observed in neighboring degree programs, when compared with the average prevalence of female applicants in the entire system. This relationship This relationship is positive, significant up to two links, and holds for both Portugal and Chile. Positive autocorrelations, up to two neighbours, are also found, although not so strong, for Application Scores (Figure \ref{figure2}g-i) and Demand-Supply ratio (Figure \ref{figure2}j-l), in both countries. 

\begin{figure*}[!t]
    \centering 
    \includegraphics[width=1.0\textwidth]{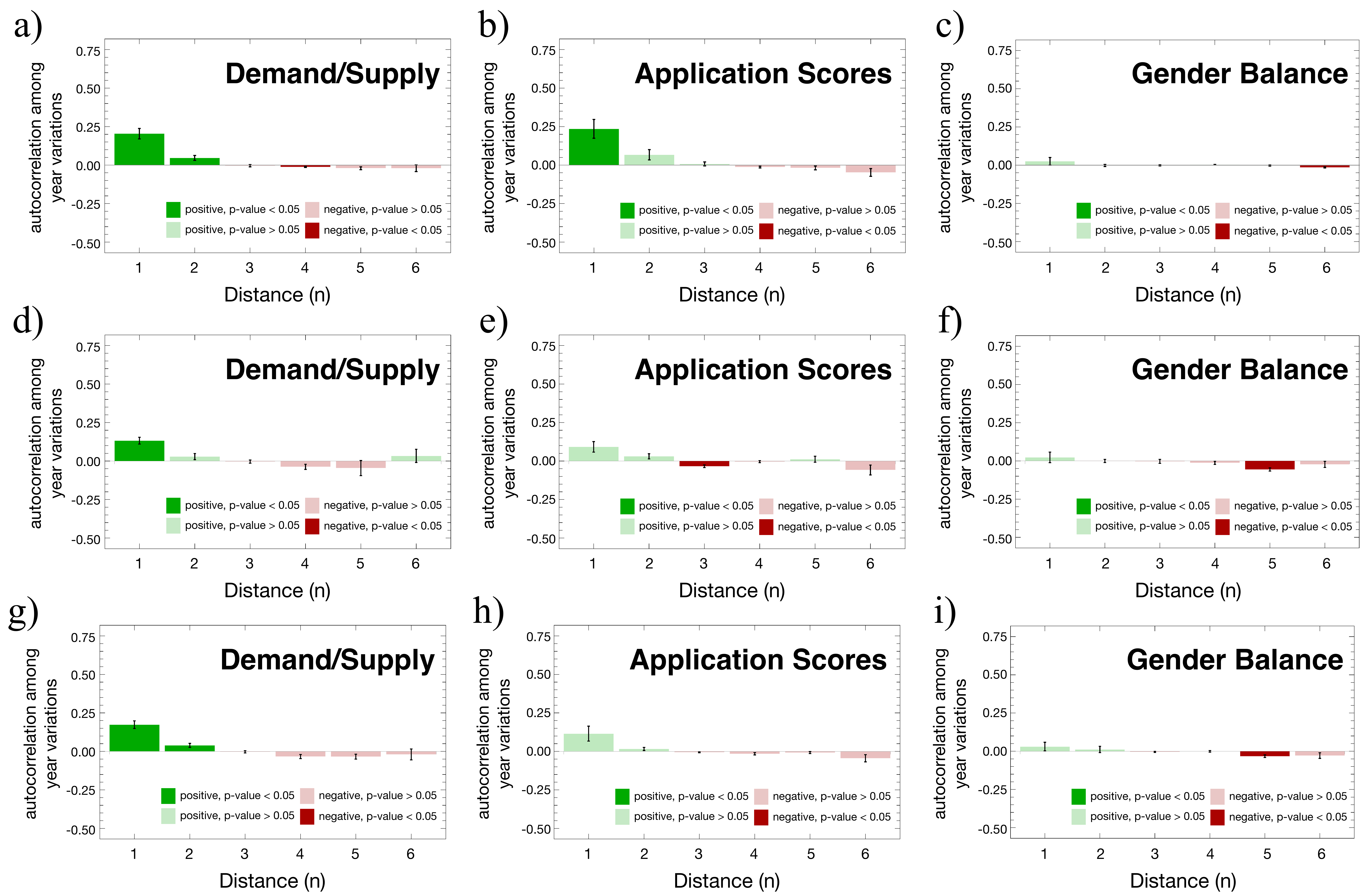}
    \caption{Auto-correlations in respect to the time variations of different features in both the PHES (a--c), and CHES networks. Panels show the  autocorrelations computed for the temporal variations of demand-supply ratio (a,d,g), application scores (b,e,h), and gender balance levels (c,f,i).
    }
    \label{figure32}
\end{figure*}

Due to data availability, autocorrelation patterns for Unemployment levels (Figure \ref{figure2}m) and Retention Rates (Figure \ref{figure2}q) are calculated for the PHES only. Both show similar behavioural patterns as in the previous features, although the positive relationship in unemployment levels extends to three-links of distance instead of two. Again, due to data constraints, the Student Mobility feature is only analyzed for the CHES (Figure \ref{figure2}n-o). Contrary to the other features, the positive relationship observed in the Geographical Mobility vanishes quickly and becomes negative/zero between degree programs separated by more than one links. Two possible explanations for the lack of a positive autocorrelation away from the first neighbors can be: 1) most applicants assign a small weight to distance as a factor in the choice of a degree program, and 2) the majority of applicants has a tendency to apply to degree programs that minimize the distance to their local of origin. Although previous research seems to support the second hypothesis \cite{sa2006does,lourencco2018spatial,van2013determinants,flannery2014they,suhonen2014field,walsh2015geographic}, a more in-depth future analysis is needed to answer this question conclusively. 

In sum, and with the exception of Geographical Mobility, all features exhibit positive autocorrelations that extend up to two/three links of separation. The Higher Education Space captures information embedded in the interplay between degree programs, which is revealed by studying the preference patterns of applicants. These results are a natural outcome of all the information applicants’ carry at the moment of their choices \cite{surowiecki2005wisdom} (i.e., either contextual information used in the decision making or inherent characteristics of applicants), which in turn modulates the topology of the HES. 

\subsection{Temporal Variations in Features}
\label{s:33}

In the previous section, we have shown \emph{How} and \emph{Where} certain degree programs are positively correlated, in several features, as a function of the network distance between them. In this section, we examine how temporal changes in these features can spillover throughout the HES. By understanding the \emph{When} of the autocorrelations patterns, it is possible, for instance, to perceive how external shocks propagate through the system. As an example, we take the particular case of the building sector in Portugal -- one of the most affected by the financial crisis that hit the country between 2010 and 2014 (a crisis that was preceded by a downward path since the beginning of the millennium and the global financial crisis of 2008 \cite{pereira2015portugal}). 

Figure \ref{figure3}a--b shows, for the PHES, the temporal variation in the demand-supply ratio for Civil Engineering (a) and Architecture (b) between 2008 and 2015. Also shown (light gray) are the temporal variations of their closest direct neighbors in the Higher Education Space network (averaged is highlighted in red).  After the economic and financial crisis, the construction industry was one of the most negatively affected \cite{pedroso2014portugal, vansteenkiste2017did}. \emph{A priori}  (without knowing the structure of the network), one could expect that both Civil Engineering and Architecture would suffer a similar impact on their demand-supply ratio given their close market relationship. However, a closer inspection of Figure \ref{figure3}a--b shows that the negative impact on the demand for Civil Engineering is not observed for Architecture. More importantly, in both cases, the variations are consistent with the average behavior of the nearest connected degree programs (temporal spillovers). This confirms and reinforces the above finding where both belong to two different clusters (architecture being closest to degree programs in Arts and Humanities than to Engineering), \emph{c.f.} Figure \ref{figure1} above. 

The spatial autocorrelation patterns, concerning the temporal variations of features, help to explain how the observed changes that affect entire regions of the network in different ways and in different time periods. For example, a clearly discernible pattern in Figure \ref{figure3}c--d reveals that variations in the demand-supply ratio reversed from one part of the network to the other in two distinct time periods (2010/11 -- Fig. \ref{figure3}c and 2014/15 -- Fig. \ref{figure3}d). These \textit{temporal spillovers} are confirmed by the autocorrelation patterns of the yearly time variations of each feature, over all degree programs in the PHES (Figure \ref{figure32}a--b). There are positive effects in time that remain up to two links of separation in the Demand-Supply Ratio and Application Scores, suggesting that, not only these two features changed over time (thus reacting to conjuntural changes) but also that those changes spillover to their neighbors.

However, we do not find autocorrelation patterns among the temporal variations for all features. Certain features, such as the demand-supply ratio (\ref{figure32} a, d, and g) and application scores (\ref{figure32} b, e, and h), show a synchronous variation over time, suggesting that it responds to  contextual changes. On the other hand, gender balance (\ref{figure32} c, f, and i) do not change over time, suggesting that it is likely to respond to more long-term structural changes, e.g., cultural mechanisms, and other socio-economic factors. In the CHES although not all autocorrelation coefficients show a statistically significant pattern, results lead to similar conclusions (see Figure \ref{figure32}d--i). 


\subsection{Measuring Unemployment Similarity}
Thus far, we have identified several prevailing autocorrelation patterns both in the spatial distribution of features but also in their temporal variations. However, at this point, it is not clear what explains the distinctive behaviour of a degree program in any given characterizing feature. For example, is the unemployment level of a degree program better explained by the connections it has in the HES or by other degree programs with similar features? 

To explore this question we compare the difference in unemployment levels in a treatment group (pairs of degree programs that are connected in the HES) against several control groups (with similar behaviours in one or more features but that are not connected). To generate the control groups, we sample, for each pair in the treatment group, a second pair with equivalent levels of similarity in the available features, namely 1) gender level, 2) application scores, 3) demand-supply levels and 4) all three features combined. In addition, we built a 1) random control group where pairs of nodes are taken at random disregarding any similarity) and 2) a control group with degree programs of the same ISCED education field.

\label{s:34}
\begin{figure}[!t]
\centering 
\includegraphics[width=1.00\linewidth]{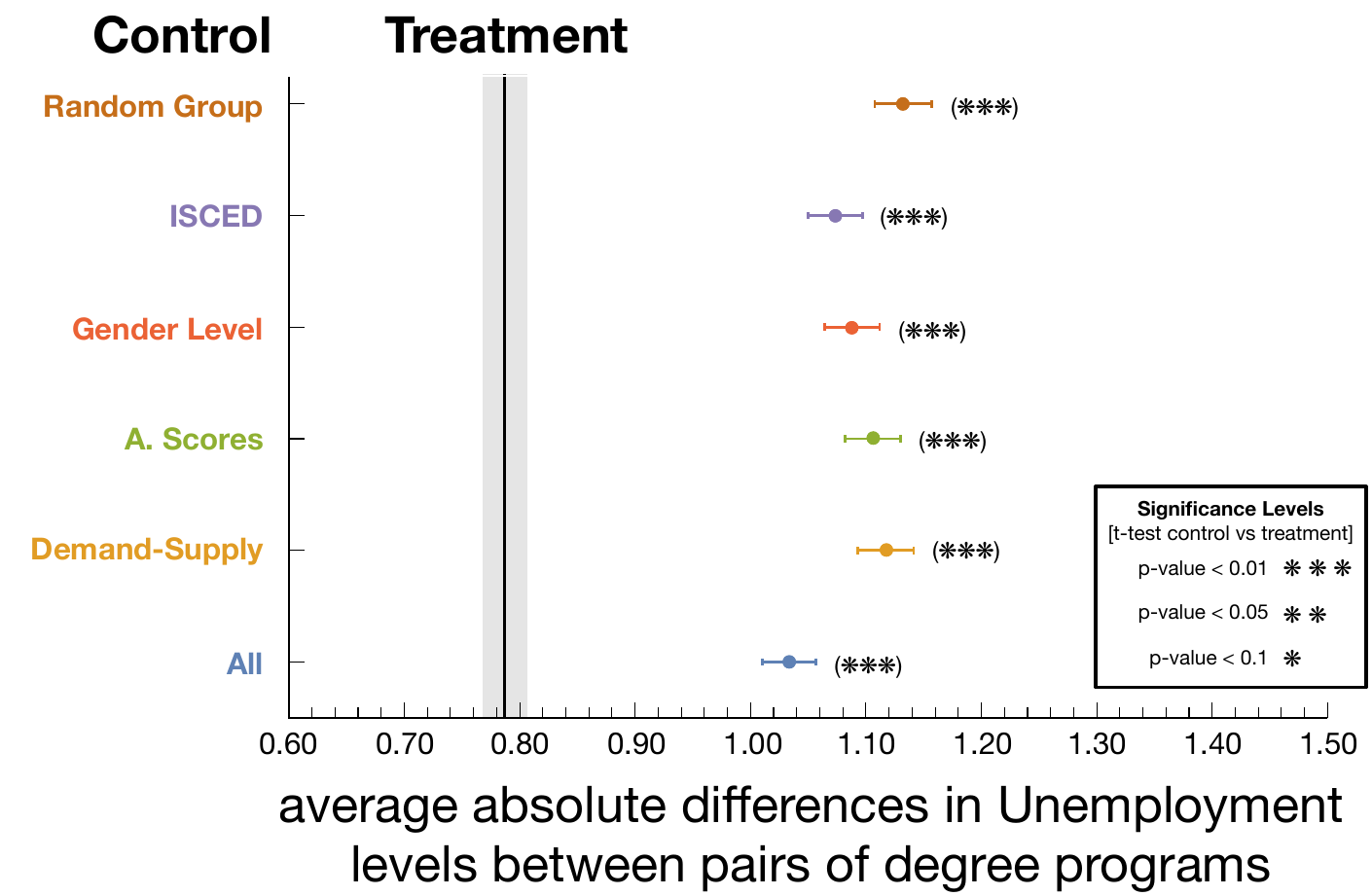}
\caption{Comparison between the absolute differences in unemployment levels of pairs of degree programs in a treatment group (black vertical line) against different control groups (horizontal). The treatment group corresponds to 1177 connected pairs of degree programs in the Higher Education Space. Each control group (of the same size as the treatment group) is a set of pairs of degree programs matched through the \textit{propensity score matching} \cite{pearl2003causality} with the pairs in the treatment group. Similarities measure the Euclidean distance among pairs of degree programs in different control groups: random (dark orange), education fields of the ISCED classification (purple), gender (red), application scores (green), demand-supply ratio (orange) and all the last three features (blue). Error bars in the control groups indicate the standard error in the estimation of the averages therein and the shaded area is the standard error for the treatment group. Statistical significance was measured by a t-test between the treatment and control averages -- all differences are significant with p-values $< 0.01$.}
\label{paper2_figure4}
\end{figure}

In Figure~\ref{paper2_figure4} rows show the average of the absolute difference in unemployment levels between pairs of degree programs for each control group. In all cases, the differences are smaller for the treatment group (vertical black line) when compared to the control groups (all differences are statistically significant -- t-test between the averages of the two groups with p-value $<0.001$). These findings support the hypothesis that the HES represents a similarity mapping between degree programs, as perceived by the applicants to higher education, that is not possible to access by estimating similarities using traditional features alone (e.g. gender, application scores or demand-supply).

We should note that nodes in these networks do not incorporate any information about the institutions. These specificities can potentially change the results of the current model, especially in those cases where factors, such as the prestige of higher education institutions, the societal value of degree programs (e.g. medicine), and the relative location of institutions to their recruitment base can greatly impact the applicants' choices \cite{simoes2010applying} and consequently, the structural organization of the higher education space.  

\section{Conclusions}
\label{conclusions}

The ever growing worldwide complexity ensuing from technological, social, cultural and economic  changes demand the design of highly effective governance instruments that can support management and policy development of higher education systems. This design requires novel data-driven approaches \cite{daniel2015big, baker2014educational} that are able to capture the complex interplay between existing elements of the system and report new, comprehensive and reliable information about its functioning. 

Here, we examined the potentials of exploring the higher education system through the lens of network science by looking at the applicants' conjoint choices and the emergent organizational structures. The rationale behind this approach originates from the assumption that students are not isolated beings when choosing their educational paths. The underlying intricacies of their choices are reflected not only on their individual decisions but also on society's organizational structure as captured and materialized in higher education systems and more specifically in the inter-dependencies among degree programs as viewed from the student's point of view.   

By leveraging the information carried by the applicants to higher education in Portugal and Chile at the time of their application we have derived wider organizing principles common to both systems. We show that the Higher Education Space (HES) is sparse, highly ordered, modular and able to capture multi-factorial information about the applicants' choices.

The HES reveals the existence of autocorrelation patterns among many features describing degree programs -- gender balance, application scores, unemployment, mobility, demand-supply ratios, and retention rates -- that stem from the aggregated characteristics of applicants and/or enrolled students. By construct, the methodology is blind to the applicants individual information, and as such, serves as evidence for validating the HES's utility as a source of non-trivial information about the system. For example, it informs that degree programs that are closer in the HES tend to be more similar in regards to their features. It follows that these similarities among degree programs have a "contagious" effect between their closest neighbours. These spatial and temporal spillovers are identified in features that reflect conjectural changes (application scores and the demand-supply). On features that reflect structural changes, as gender balance, only spatial spillovers are identified. 

Moreover, the connectivity structure of the HES offers a larger explanatory power to certain features, such as unemployment levels, than a proximity mapping using other traditional variables. This implies an important take away for applicants, as unemployment is prevalent in full regions (\textit{i.e.}, sets of interconnected degree programs), thus exhibiting above-average unemployment, which can later manifest as a job mobility cost for graduates.

As Baker \cite{baker2018understanding} stresses, perception mismatches between students or applicants and educators or decision-makers need to be taken into consideration when developing new policies. In that sense, here we proposed a network driven classification of degree programs that can serve as a complement to the ISCED classification. In our classification degree programs are grouped according to the applicants' perspectives, not to their curricular content. The HES stems from a much richer and multi-factorial decision-making process than the ISCED classification, reflecting how actors in the society perceive higher education.

As stated in the beginning of this work, we aimed at showing the potential of the \emph{Higher Education Space} in supporting policy development. Admittedly, much was left for future work. In this respect, we identify three main areas for future development: 1) exploring the practical and actual application of the HES in designing effective governance actions, 2) exploring the resulting topological features of the HES for a wider spectrum of countries. This can either highlight the universality of the structures identified or help us understand how distinct HES are shaped by different cultural contexts and perspectives. Finally, 3) in countries where application systems are not governed by a central body -- such as in the USA and in Brazil -- the methodology herein can be replicated by resorting to nationwide surveys that mimic the application process in countries such as Portugal and Chile.



\section*{Acknowledgements}
CC and FLP acknowledge the financial support from the MIT Media Lab Consortia and the MIT-Masdar Instiute (USA—Reference 0002/MI/MIT/CP/11/07633/GEN/G/)  initiative. CC acknowledges the financial support from Centro de Investigación en Complejidad Social. SE acknowledges the financial support of FCT/MCTES through National funds, for project grants SFRH/BPD/1169337/2016 and by multi-annual funding of CICS.NOVA (UID/SOC/04647/2013). SE and FLP acknowledge the financial support of A3ES (2015-2017). The authors are thankful to Carlos Rodriguez-Sickert for logistic support with the Chilean data set,  Jorge Jara Ocampo for collecting data and to C\'{e}sar A. Hidalgo, Cecilia Monge, Mary Kaltenberg, Aamena Alshamsi, Tarik Roukny, V\'{i}tor V. Vasconcelos, Diana Orghian, Alberto Amaral, Madalena Fonseca, Gustavo Castro-Dominguez, Carla Sá, the Collective Learning Group at the MIT Media Lab, Laszlo Barabási and the Center for Network Science Research at Northeastern University for the helpful insights and discussions.

\renewcommand\thefigure{\thesection\arabic{figure}}
\appendix
\setcounter{figure}{0}
\section{Data Characterization}
\label{a:datadescription}

\begin{figure*}[!t]
 \centering 
  \includegraphics[width=1.0\textwidth]{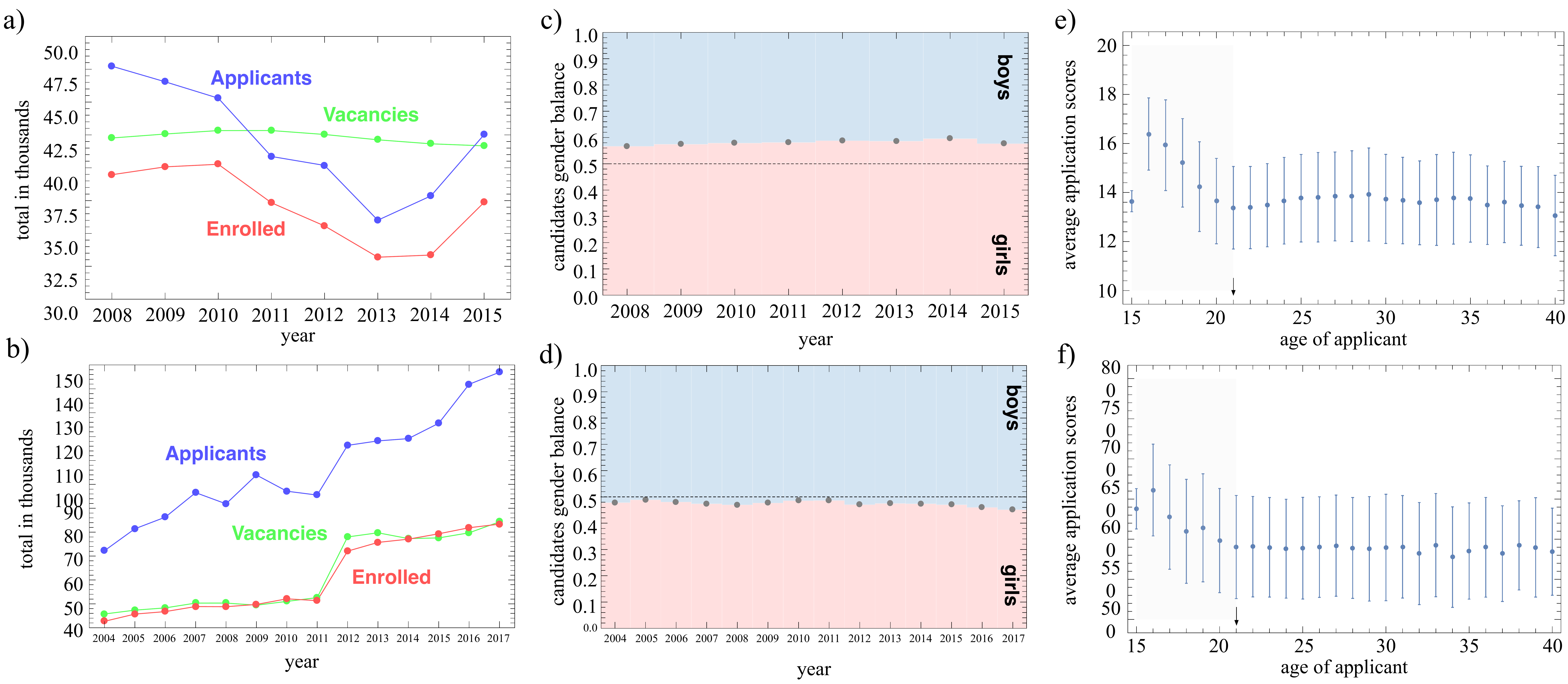}
\caption{Descriptive information of the Higher Education Systems of Portugal (upper row) and Chile (bottom row). Panels a and b show the time evolution of the number of Candidates (blue), number of open positions or vacancies (Green), and number of enrolled students (red) in the Portuguese (a) and Chilean (b) Higher Education System. Panels c and d show the time evolution of the gender balance among Candidates of the Portuguese (a) and Chilean (b) Higher Education System. Blue represents the fraction of boys and pink the fraction of girls.
}
 \label{fig:A1}
\end{figure*}

There are three key aspects in common between the Portuguese and Chilean Higher Education Systems. First, the application process is centralized and controlled by state-governed bodies. Second, the application consists in the submission of a ranked list of preferences – up to six in Portugal and ten in Chile – that correspond to pairs Institution and Degree Program. Third, candidates are allocated to open positions – which are set prior to the competition process – by descending order according to the candidates’ preferences and scores.

The Portuguese higher education system is organized into Universities and Polytechnics, and the Public sector represented more than $70\%$ of total students’ enrollments (first year, first time for all education levels), between 2004 and 2015. Moreover, the competition and centralized application process applies only to public institutions and takes three rounds sequentially. After each round, candidates submit a new list of preferences and a new list of open positions is updated by institutions. Applications to private institutions are submitted directly to the institution but must follow the general rules legally established for accessing higher education (e.g. minimum application score).

Public Portuguese Higher Education Institutions included in this study are: Universidade dos Açores; Universidade do Algarve; Universidade de Aveiro; Universidade da Beira Interior; Universidade de Coimbra; Universidade de Évora; Universidade de Lisboa; Universidade Técnica de Lisboa; Universidade Nova de Lisboa; Universidade do Minho; Universidade do Porto; Universidade de Trás-os-Montes e Alto Douro; Universidade da Madeira; Universidade Aberta; Instituto Politécnico de Beja; Instituto Politécnico do Cávado e do Ave; Instituto Politécnico de Bragança; Instituto Politécnico de Castelo Branco; Instituto Politécnico de Coimbra; Instituto Politécnico da Guarda; Instituto Politécnico de Leiria; Instituto Politécnico de Lisboa; Instituto Politécnico de Portalegre; Instituto Politécnico do Porto; Instituto Politécnico de Santarém; Instituto Politécnico de Setúbal; Instituto Politécnico de Viana do Castelo; Instituto Politécnico de Viseu; Instituto Politécnico de Tomar; Instituto Superior de Ciências do Trabalho e da Empresa; Escola Superior de Enfermagem de Coimbra; Escola Superior de Enfermagem de Lisboa; Escola Superior de Enfermagem do Porto; Escola Superior de Enfermagem de Artur Ravara; Escola Superior de Enfermagem de Maria Fernanda Resende; Escola Superior de Enfermagem de Francisco Gentil; Escola Superior de Enfermagem de Calouste Gulbenkian de Lisboa; Escola Náutica Infante D. Henrique; Escola Superior de Hotelaria e Turismo do Estoril.

The Chilean higher education system is organized into Universities, Professional Institutes, and Technical Schooling. Universities are classified into ‘Traditional’ or Private. ‘Traditional’ universities, created before de 1981 educational reform, include public universities belonging to the CRUCH\footnote{CRUCH stands for \emph{Consejo de Rectores de las Universidades Chilenas}} and private universities with state funding. The competition process, called \emph{Sistema Único de Admisión} – SUA, was implemented in 2003 and is managed by the DEMRE. It started by covering the access to just 27 ‘Traditional’ universities but following an educational movement commanded by secondary school students in 2012, 9 other private universities were included. Contrary to its Portuguese counterpart, the Chilean application process happens in a single round. Until 2011, SUA represented around $44.4\%$ of total enrollments in universities and since 2012 around $67.9\%$.

The traditional/public higher education institutions in the Chilean included in this study are: Universidad de Chile; Pontificia Universidad Catolica de Chile; Universidad de Concepcion; Pontificia Universidad Catolica de Valparaiso; Universidad Tecnica Federico Santa Maria; Universidad de Santiago de Chile; Universidad Austral de Chile; Universidad Catolica del Norte; Universidad de Valparaiso; Universidad Metropolitana de Ciencias de la Educacio;Universidad Tecnologica Metropolitana; Universidad de Tarapaca; Universidad Arturo Prat; Universidad de Antofagasta; Universidad de la Serena; Universidad de Playa Ancha; Universidad de Atacama; Universidad del Bio-Bio; Universidad de La Frontera; Universidad de Los Lagos; Universidad de Magallanes; Universidad de Talca; Universidad Catolica del Maule; Universidad Catolica de La Santisima Concepcion; Universidad Catolica de Temuco; Universidad de O'higgins; Universidad de Aysen.
Private Chilean Higher Education Institutions included, starting from 2011, are: Universidad Diego Portales; Universidad Mayor; Universidad Finis Terrae; Universidad Andres Bello; Universidad Adolfo Ibañez; Universidad de Los Andes; Universidad del Desarrollo; Universidad Alberto Hurtado; Universidad Catolica Silva Henriquez.            

Figure \ref{fig:A1}a and \ref{fig:A1}b show the time evolution of the number of candidates (blue), number of open positions (green), and number of enrolled students (red) per year, both for Portugal and Chile. It is important to note that the Portuguese Higher Education system suffered an overall decrease in demand during the period of analysis, follow by a recovery (although by 2018 this number has yet failed to match 2008 values), which contrasts with the steady growth observed in Chile during the same period. The decline of the Portuguese demand can have its origin not only in demographic trends but also in a wide range of socio-economic factors, this is however out of the scope of this work.

Figure \ref{fig:A1}c and \ref{fig:A1}d shows the time evolution of the number of candidates (blue), number of open positions (green), and number of enrolled students (red) per year, both for Portugal and Chile. It is important to note that the Portuguese Higher Education system suffered an overall decrease in demand during the period of analysis, follow by a recovery (although by 2018 this number has yet failed to match 2008 values), which contrasts with the steady growth observed in Chile during the same period. The decline of the Portuguese demand can have its origin not only in demographic trends but also in a wide range of socio-economic factors, this is however out of the scope of this work.

\begin{figure*}[!t]
 \centering 
  \includegraphics[width=1.0\textwidth]{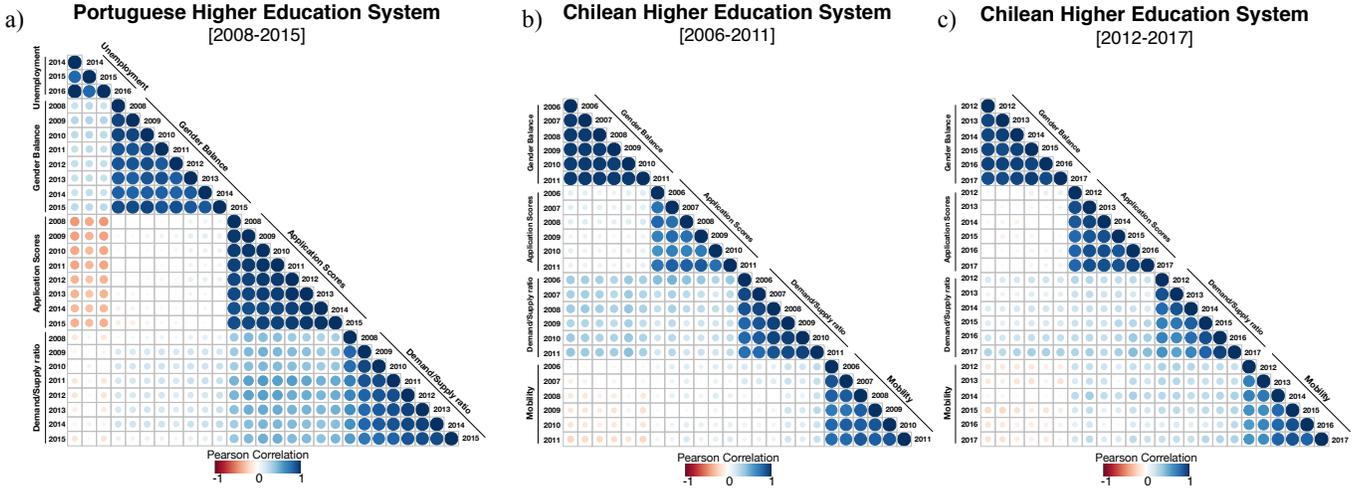}
\caption{Correlations among the standardized features of degree programs for the Portuguese (a) and Chilean (b and c) Higher Education System. The features correspond to the candidates (first option of the preferences) and enrolled students characteristics aggregated by degree program and per year, plus the unemployment data per degree program. Color indicates the Pearson correlation, red is negative and blue positive. Size of each disk indicates the magnitude of the correlation.}
 \label{fig:A2}
\end{figure*}

Figure \ref{fig:A1}e and \ref{fig:A1}f shows the average application score of candidates aggregated by age to both the Portuguese (e) and Chilean (f) Higher Education Systems. For both cases, the average scores peak at the age of sixteen years old and tend to decrease with older candidates, apart from fifteen years old that tend to have lower average scores when compared with the former. In both cases, the majority of the (above $90\%$) candidates are twenty-one or younger, and roughly $80\%$ are eighteen or nineteen years old.

Figure \ref{fig:A2} show per year the correlations among the standardized features that mark degree programs. These features include the aggregated characteristics of enrolled students (gender, application scores, mobility), the first option of candidates (demand over supply ratio), but also the output variables such as the Unemployment level. In all cases, correlations are very strong between the same features across the years and for that reason we have opted to discuss, in the main manuscript, the results for a single year for all networks (typically the last year available). Interestingly, correlation patterns between different features in the data, some of which more ‘common sense’ than others, include the following: Unemployment shows a small but positive correlation with increasing proportion of girls and a negative correlation with application scores; for Chile, increasing proportions of girls are positively correlated with demand and negatively correlated with mobility; in both cases demand is positively correlated with application scores.
\setcounter{figure}{0}
\section{Data Cleaning}
\label{a:datacleaning}

In this appendix we discuss the data transformations procedures in order to clean and filter the original raw data. We divide this Appendix in two subsection as the data cleaning process had differences between the two systems under study.

\subsection{Portuguese Higher Education System Data Set}
The level of organization of the raw data facilitated, greatly, the cleaning and filtering tasks.
Besides a name, degree programs are encoded by a unique 4-digit ID code, and also have a 3-digit ISCED classification code associated with it. Degree programs are divided in several types, which are important to mention. First, there are 3-year (BA) and 5-year degree programs (BA+MA), both of which are available as options for candidates applying for the first time to Higher Education but offer different output degrees. Second, degree programs can be taught in different regimes. For instance, they can be taught in Portuguese or in a foreign language (e.g. English or French), and they can also be taught during the daytime (normal) or in a nocturnal (special) regime. 
In some cases, degree programs with the same name, can have different IDs to differentiate between these different types/scenarios. In order to clean and disambiguate degree programs, we have done the following steps:

Other important manipulations include discarding all applicants older than 21 years old, in order to exclude applicants that entered in the Higher Education System via special programs. Additionally, we have only considered the first round of the Portuguese application process in all subsequent analyses. 

\subsection{Chilean Higher Education System Data Set}
Data from Chile comes from multiple sources and, as such, it was not disambiguated at the same level as in the Portuguese data set. One major issue is that there is a unique ID for each pair degree programs and institution. For instance, the degree program in Physics will have different identification IDs, one per institution. A second issue is that we only have information on the first two levels of the ISCED classification. To disambiguate this problem we applied the following steps: 1) Discarded all degree programs that are not taught during in Portuguese during the normal regime (day time); 2) Aggregate degree programs with the same name but different IDs.

As in the Portuguese dataset, here we have also discarded all applicants older than 21 years old.

\setcounter{figure}{0}
\section{Higher Education Construction and Network Science Methods}
\label{a:networks}

\begin{figure*}[!t]
 \centering 
\includegraphics[width=\textwidth]{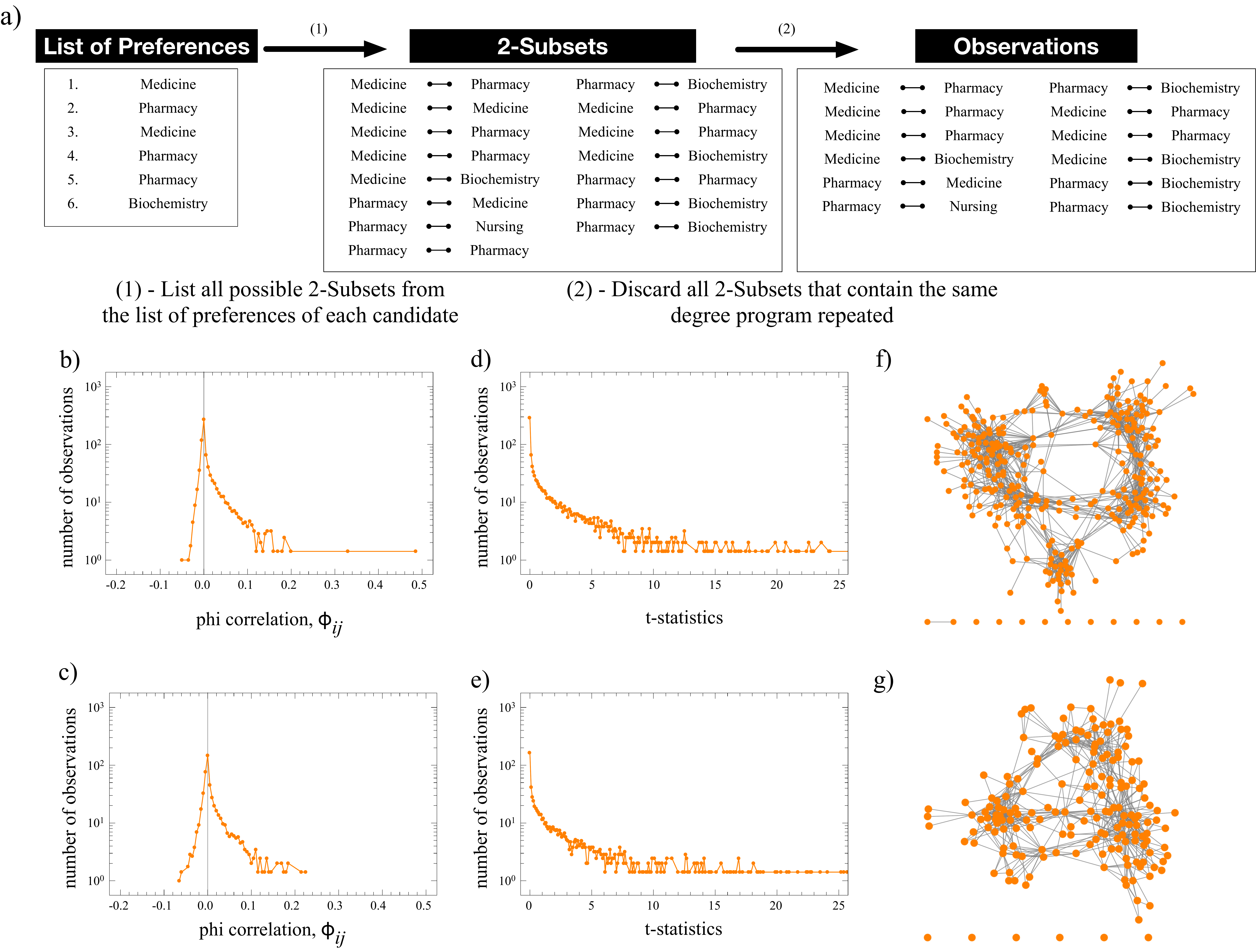}
\caption{a) Illustration of the procedure used to extract observations from the initial list of preferences of each candidate. From each candidate’s preference list, we construct a list of all possible 2-Subsets. From the latter, we discard all sets that contain the same degree program repeated. The final list corresponds to a list of pairs of degree programs.
b--g) Depiction of the steps conducted to generate the PHES and CHES networks. Step 1 (b,c), observations are collected, Step 2 (d,e) we compute the Phi correlation discarding all edges that exhibit negative values, next we compute the T-Statistics discarding all edges that are non-significant with a p-value < 0.05. Step 3 (f,g), we discard all nodes that are unconnected from the giant component.}
 \label{a:cleaningscheme}
\end{figure*}

The initial data set comprises of a list of preferences for each unique candidate. From here, a list of pairs of co-occurrences is generated among the preferences of each candidate. We do this by constructing, for each candidate, a list of all possible 2-Subsets, discarding all 2-Subsets that have repeated entries. This process is graphically depicted in Figure \ref{a:cleaningscheme}a, we refer to the final list of 2-Subsets as the observations.

After obtaining a list of all observations, we build a symmetric matrix that counts all co-occurrences of each pair of degree programs in the preference list of candidates, followed by the following steps: 1) ignore self-connections, 2) calculated correlations -- consider only positive values, 3) compute t-statistics and 4) select all significant edges (p-value $< 0.05$) and 5) discard all nodes that are unconnected from the giant component of the network. This process is illustrated in Figure \ref{a:cleaningscheme}c--g.

\subsection{Network Science Methods}

A network is a system comprised by a set of nodes/vertices and a set of links/edges. Links represent a pair of nodes that are connected. In the particular case of this manuscript, nodes abstract degree programs, and links represent a statistically significant co-occurrence relationship between a pairs of degree programs. The connectivity $k_i$ of a degree program, $i$, measures the number of degree programs it is connect to by a link. Figure \ref{figure1} shows the graphical representation of the Higher Education Space network. Alternatively, one can represent the network through an adjacency matrix, $A$, where each entry ($a_{ij}$) is one if there is a link that connects degree programs $i$ and $j$, or zero otherwise. Given the nature of the Higher Education Space, the adjacency matrix is symmetric, and all elements of the diagonal are zero.
The degree distribution ($D(k)$) indicates the fraction of nodes in a network with degree $k$. Hence, the average degree ($\langle k \rangle$) of a network corresponds to  $\langle k \rangle = \sum_k k D(k)$, while the degree variance ($\text{var}(k)$) is given by $\langle k \rangle^2 - \langle k^2 \rangle$.

The \textit{Average Path Length} (APL) measures the average shortest distance, measure in links, between any two nodes the network. That is, the minimum number of links a random walker would have to transverse if going from one node to the other \cite{Barabasi2016}, averaged by all possible pairs of nodes. The APL can be formally computed as
\begin{equation}
    APL=\frac{1}{Z(Z-1) \sum_{i\neq j} d(i,j)}
\end{equation}
where $Z$ is the number of nodes in the network, and $d(i,j)$ the shortest distance between nodes $i$ and $j$ measured in number of links. The \textit{Cluster Coefficient} $C$ measures the average fraction of triangular motifs a node participates in over the total number of possible triangles \cite{newman2003structure}. Formally, it can be computed as 
\begin{equation}
    C=\frac{1}{Z}\sum_i \frac{\lambda_i}{\Lambda_i}    
\end{equation}
where $\lambda_1$ is the number of triangles that involve $i$, and $\Lambda_i$ is the number of triples that involve $i$, that is, the number of sets of three nodes with two edges (open triangle).

The \textit{Modularity} (Q) measures the quality of a particular partition (\textit{i.e.}, groups of nodes) of a graph by estimating how many links between elements of the same group are captured by a given partition when compared with a random and uncorrelated network with the same connectivity distribution \cite{newman2006modularity}. Formally, Q is computed as
\begin{equation}
    Q=\frac{1}{2m}\sum_{i,j}\left[ a_{ij} - \frac{k_i k_j}{2m} \delta (c_i,c_j) \right]
\end{equation}
where $m$ is the total number of links in the network, $a_ij$ equals one if $i$  and $j$ are connected and zero otherwise, $k_i$ is the number of links where node $i$ participates, its degree, $\delta (X,Y)$ equals one if $X=Y$  and it is zero otherwise, and $c_i$ is the community/partition that node $i$ belongs.
\setcounter{figure}{0}
\section{Feature Assortment on the Higher Education Space and Autocorrelations}
\label{a:features}

In this appendix we expand the discussion of the revealed spatial correlations. We start by showing in more detail how the autocorrelation coefficients have been obtained. We follow with a illustrative examples of the spatial assortment of the Application Scores, Unemployment, Demand and Mobility over the Higher Education Spaces of Portugal and Chile.

\begin{figure*}[!t]
 \centering 
\includegraphics[width=\textwidth]{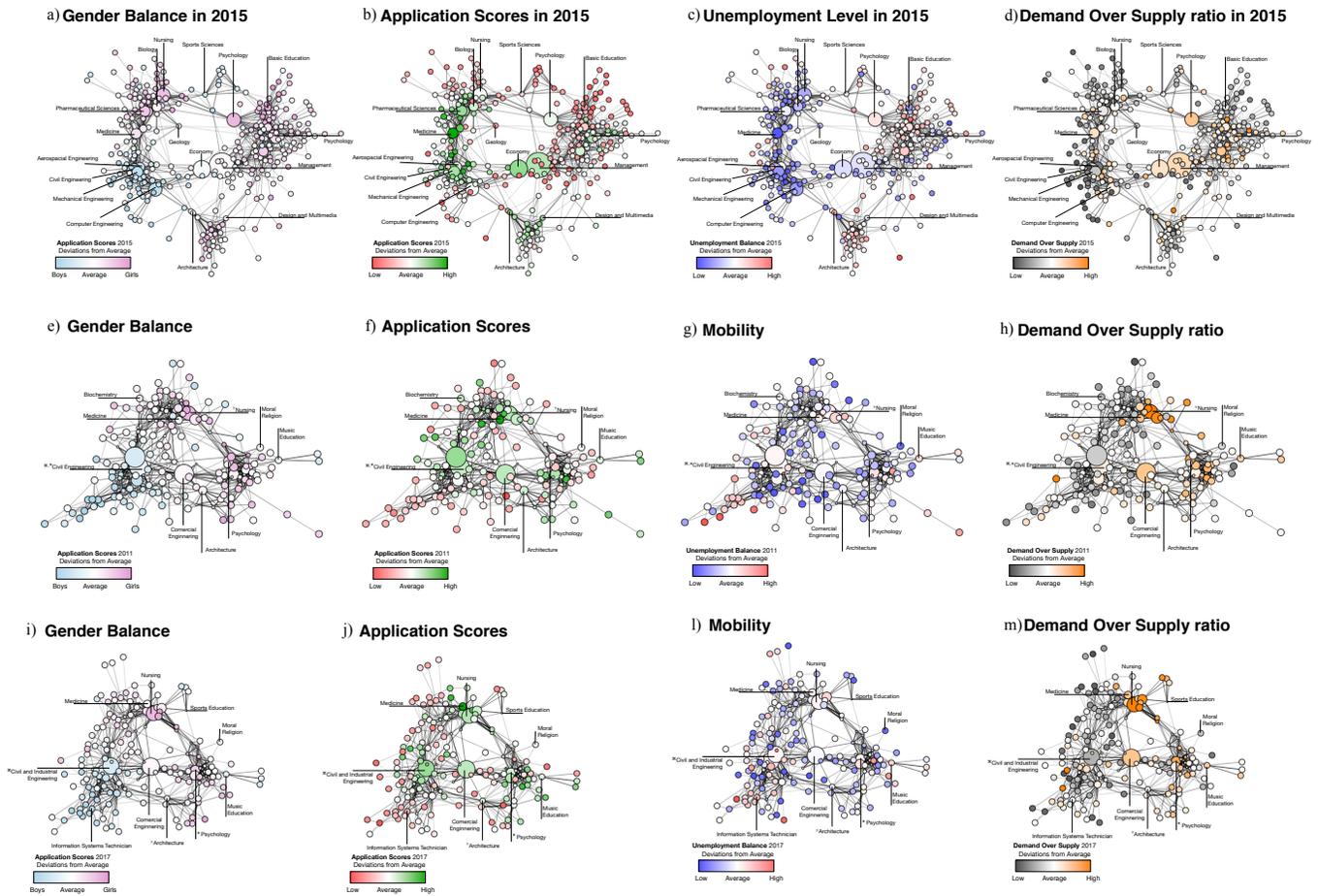}
\caption{a) Illustration of the procedure used to extract observations from the initial list of preferences of each candidate. From each candidate’s preference list, we construct a list of all possible 2-Subsets. From the latter, we discard all sets that contain the same degree program repeated. The final list corresponds to a list of pairs of degree programs.
b--g) Depiction of the steps conducted to generate the PHES and CHES networks. Step 1 (b,c), observations are collected, Step 2 (d,e) we compute the Phi correlation discarding all edges that exhibit negative values, next we compute the T-Statistics discarding all edges that are non-significant with a p-value < 0.05. Step 3 (f,g), we discard all nodes that are unconnected from the giant component.}
 \label{a:spatialdistribution}
\end{figure*}

\begin{figure*}[!t]
 \centering 
\includegraphics[width=\textwidth]{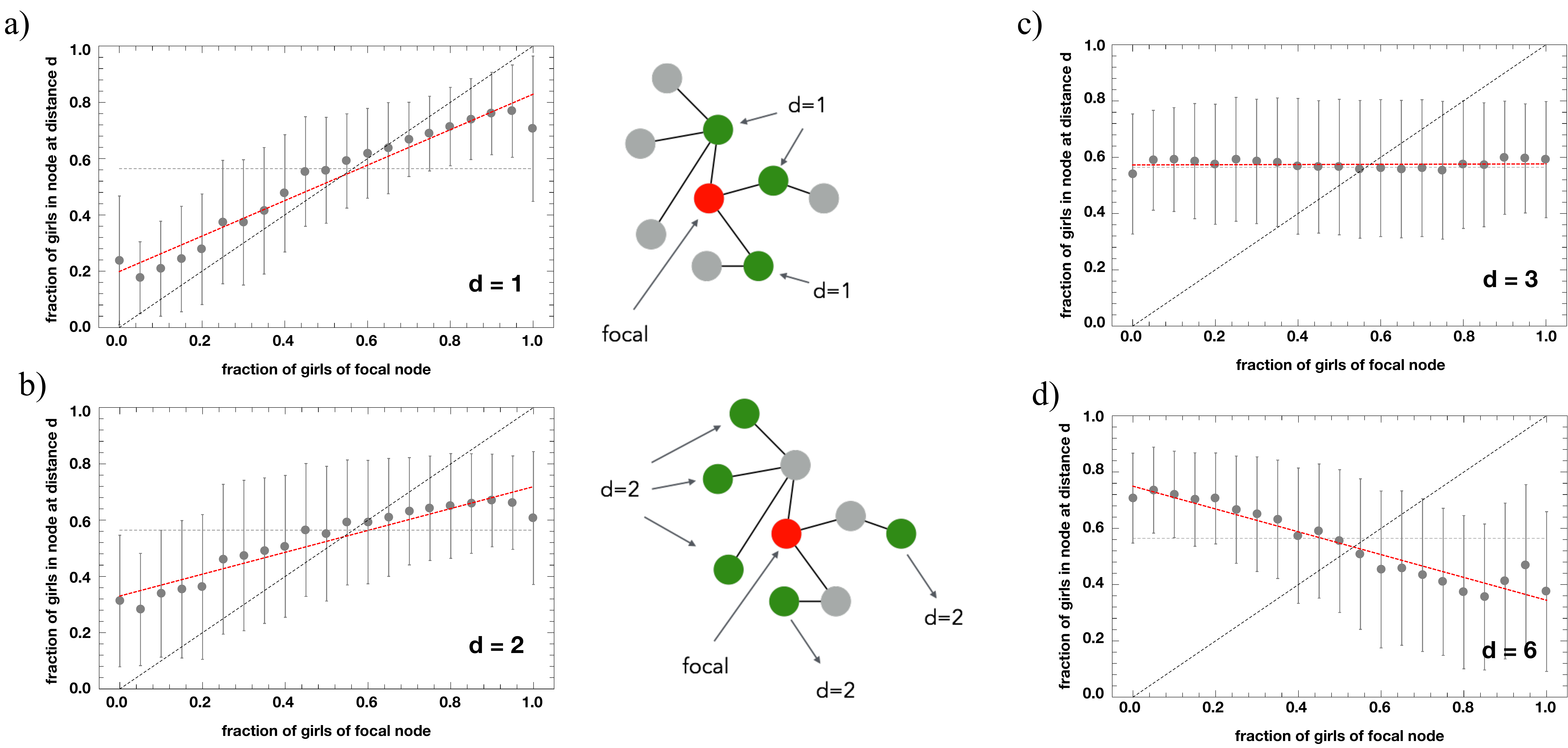}
\caption{Spatial correlation method in the higher education space. A) Regressing the fraction of girls of all first-neighbors of the focal node (red node) on the fraction of girls of focal nodes. B)  Regressing the fraction of girls of all second-neighbors of the focal node (red node) on the fraction of girls of focal nodes. C) Regressing the fraction of girls of all third-neighbors of the focal node (red node) on the fraction of girls of focal nodes. D) Regressing the fraction of girls of all sixth-neighbors of the focal node (red node) on the fraction of girls of focal nodes.}
 \label{a:autocorrelations}
\end{figure*}

Figure \ref{a:spatialdistribution} shows how different degree program attributes such as, gender balance, average application scores, demand/supply ratio, and unemployment level, are distributed over the Portuguese (a--d) and Chilean (e--n) higher education spaces. In order to quantify how the HES captures different features of the higher education system, we regress the average level of a degree program attribute among the k-nearest neighbors on the average level of the same  attribute for a single degree program or focal node. Figure \ref{a:autocorrelations} illustrates the method to estimate the autocorrelation coefficients. For each feature we take a focal node, then we assign the average level of the attribute of the first-neighbors to the independent variable vector and the level of the attribute of the focal node to the dependent variable vector. We repeat the same process for each node as a focal node. Finally, we repeat all for until the sixth-neighbors. For each k-level of neighbors we run a linear regression to estimate the strength of the relationship based on the linear regression coefficient.   

\bibliography{references}
\end{document}